\documentclass{nature}
\usepackage{color} 
\usepackage{graphicx}
\usepackage[]{algorithm2e}
\usepackage{listings}
\usepackage{footnote}
\usepackage[CaptionAfterwards]{fltpage}
\usepackage{ccaption}
\makesavenoteenv{tabular}
\makesavenoteenv{table}

\title{Super-resolved 3-D imaging of live cells' organelles from bright-field photon transmission micrographs}

\author{Renata Rycht\'{a}rikov\'{a},$^{1}$ Tom\'{a}\v{s} N\'{a}hl\'{i}k,$^1$ Kevin Shi,$^2$ Daria Malakhova,$^1$ Petr Mach\'{a}\v{c}ek,$^{1}$ Rebecca Smaha,$^{3}$ Jan Urban,$^1$ \& Dalibor \v{S}tys$^1$}

\begin{document}
\maketitle

\begin{affiliations}
\item Institute of Complex Systems, Faculty of Fisheries and Protection of Waters, University of South Bohemia, Z\'{a}mek 136, 373 33 Nov\'{e} Hrady, Czech Republic.
\item Princeton University, Princeton, New Jersey 08544, USA.
\item Department of Chemistry, Stanford University, Stanford, California 94305, USA.
\end{affiliations}

\begin{abstract}
Current biological and medical research is aimed at obtaining a detailed spatiotemporal map of a live cell's interior to describe and predict cell's physiological state. We present here an algorithm for complete 3-D modelling of cellular structures from a z-stack of images obtained using label-free wide-field bright-field light-transmitted microscopy. The method visualizes 3-D objects with a volume equivalent to the area of a camera pixel multiplied by the z-height. The computation is based on finding pixels of unchanged intensities between two consecutive images of an object spread function. These pixels represent strongly light-diffracting, light-absorbing, or light-emitting objects. To accomplish this, variables derived from R\'{e}nyi entropy are used to suppress camera noise. Using this algorithm, the detection limit of objects is only limited by the technical specifications of the microscope setup--we achieve the detection of objects of the size of one camera pixel.  This method allows us to obtain 3-D reconstructions of cells from bright-field microscopy images that are comparable in quality to those from electron microscopy images.
\end{abstract}


\begin{figure}
\end{figure}

\section*{Keywords} super-resolution, bright-field microscopy, light transmission, live cell imaging, 3D

\section*{Highlights} The choice of method for microscopic observation is in most cases limited by the possibility of sample preparation. This is particularly significant in biology of live cells, where the sample is sensitive to any sub-optimal growth conditions. It is known for many years that minute diffracting objects of 25 nm in diameter may be observed inside the living cell at high light intensities. This article describes a mathematical and technical method which utilize an ordinary bright-field microscope to obtain localization of objects inside a live cell up to the voxel of 34 $\times$ 34 $\times$ 130 nm$^3$. We believe that this approach may constitute a breakthrough in the microscopy of diffracting nano-objects in general and live cells in particular.

\section*{INTRODUCTION}




\indent \indent Bright-field microscopy is a classical method, favored for its convenience and ability to observe the physiology and morphology of unlabelled living cells and tissues. It avoids potentially complicated sample preparation procedures and visual artifacts due to complex optical paths and, in addition, is non-destructive. However, the main issue that hinders the segmentation and analysis of bright-field microscopy images\cite{alexandrov,vissapragada,candia,selinummi,mohamadlou,wu,zaritsky,buggenthin} is the low contrast of structures in the focal plane caused by distortions from an object spread function (OSF), which is unknown for most objects. These distortions are particularly relevant in a biological context, as biological specimens are significantly thicker than the depth-of-field of typical bright-field microscope lenses\cite{turner} and also have particular physicochemical properties that lead to optical inhomogeneities and further complicate the OSF. Its analysis is in addition complicated by the dynamic nature of living cells, which causes spatiotemporal changes in the image. Finally, the discretization performed during image capture may also produce inaccuracies. The resulting standard bright-field microscopy image represents multiple processes and exhibits a multifractal character.


These issues impose several constraints on the type of algorithm and microscope appropriate for this task:

\begin{enumerate}
\item It is necessary to obtain the most real and natural images possible in order to discover the spectral properties of a cell's spread function. This can be carried out using a high-resolution camera equipped with an image sensor overlaid with a Bayer filter, capturing RAW files in a higher-bit colour depth and processing them using an non-interpolating algorithm.\cite{rychtarikova,tkacik} Precise microscope mechanics should ensure the smallest possible movement along the z-axis.
\item The analytical method must be able to recognize spontaneous and random processes that underlie self-organization and multifractality.\cite{stysiwsos} Extracting the information from an image using R\'{e}nyi entropy\cite{renyi} parametrized by $\alpha$ ($\alpha \geq 0$ and $\alpha \neq 1$)  serves as an appropriate basis for this task. 
\item 
The method must be sensitive to diffraction, which is the main interactive process between light and cellular structures. Properties of the light wavefront that arises from  diffraction and is projected at the objective lenses are described in full by Mie scattering theory.\cite{Mie1908} Under the condition that the size of a particle is much larger than the wavelength of light, ray tracing techniques (geometry optics) provide a sufficient model for the characterization of the shape of the particle. Then, the behaviour of light at the interface of the strong diffracting object can be described by the phenomenon of total external light reflection and diffraction (\textbf{Supplementary Fig. 1b}). 
\item 
The method must recognize the focus of the cell in its spread function. According to the Extended Nijboer-Zernike (ENZ) theory,\cite{braat,zernike,nijboer} the focus of a fluorescent and light-diffracting object is located at the position of the highest and lowest energy density, respectively (\textbf{Supplementary Fig. 1a}).
\end{enumerate}

Here, we demonstrate a novel mathematical approach to reach superresolution in bright-field microscopy. This method, validated using atomic force microscopy, was applied to 3-D reconstructions and spectral and dynamic analysis of organelles and OSFs from z-stacks of bright-field microscopy images of live mammalian cells.

\section*{RESULTS}

\indent \indent The method is demonstrated on two cells of MG-63 human osteosarcoma (labelled \textbf{a} and \textbf{b}) from different cultivation batches and a cell of L929 mouse adipose tissue; the z-stacks of 12-bit bright-field microscopic RAW files were collected with an average z-step of 119, 150, and 158 nm, respectively. The detailed scanning conditions are described in \textbf{Table 1}. The z-stacks underwent image pre-processing such as vertical image registration (the MG63-\textbf{a} cell) and the removal of defective (dead and hot) camera pixels (the MG63-\textbf{b} and L929 cells) to avoid image defects, which, in addition, demonstrates the robustness of the method.

The overall preview of the image processing of the z-stack of the input data---12-bit RAW files with a cell of interest and background---with respect to the items mentioned above is shown in \textbf{Fig. 1a} and discussed in detail in the following sections.

\begin{figure}
  \centering
  \includegraphics[width=\textwidth]{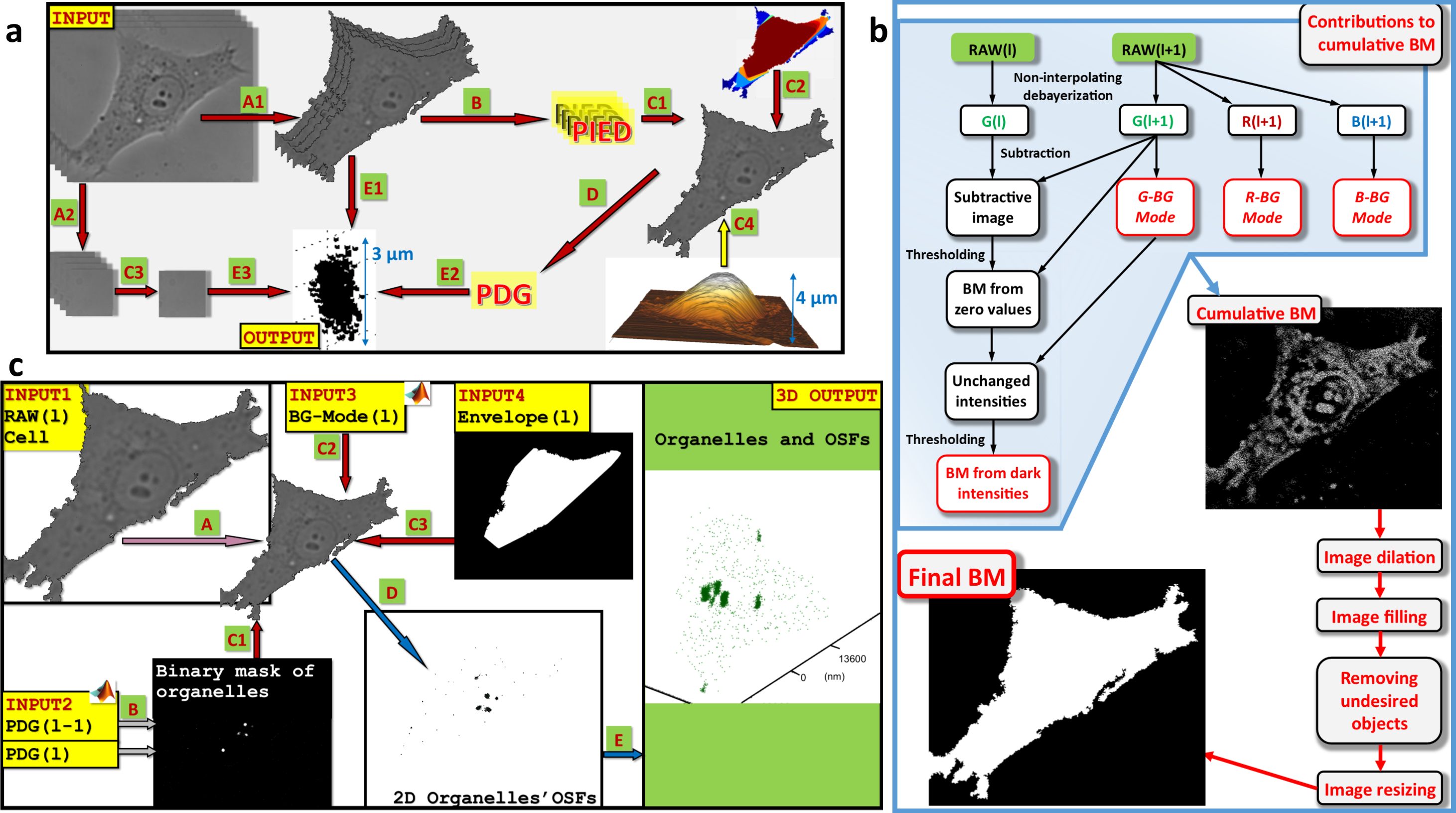}
  \caption{Scheme of the algorithm for 3-D reconstruction of organelles inside a live cell from bright-field photon microscopy (illustrated on stable homogenous diffracting organelles inside the MG63-\textbf{a} cell). (\textbf{a}) Total overview of the algorithm.  \textbf{A1} -- 2-D cell segmentation from the original input z-stack images (in 12-bit RAW files, \textbf{Algorithm 1}); \textbf{B} -- calculating $\Xi_{\alpha,c}$ (PIED) spectra, where $c$ is a colour channel, for each z-stack image; \textbf{C1} -- selection of the focal region of the z-stack according to $\Xi_{\alpha}$-spectra (\textbf{Algorithm 2}); \textbf{C2} -- calculation of the cell topography (\textbf{Algorithm 3}); \textbf{C4} -- comparison of the z-stack region with the AFM profile of the cell (\textbf{Supplementary Fig. 4}); \textbf{D} -- calculating $\omega_{\alpha,l,x,y,c}$ (PDG) values from two consecutive z-stack images; \textbf{A2} -- characterization of the background of each image as a mode of its R-, G-, and B-intensity histograms, respectively;}
  \label{Fig1.png}
\end{figure}
\begin{figure}
  \contcaption{\textbf{C3} -- selection of background values in the focal plane (complementary to the image of the cell in the focal region); \textbf{E1}, \textbf{E2}, \textbf{E3} -- 3-D organelle segmentation and reconstruction (the output of \textbf{Algorithm 4}) from the focal region of the cell, including its $\omega_{\alpha,l,x,y,c}$-images and background values. (\textbf{b}) Detailed scheme of the cell segmentation (\textbf{A1} process in panel \textbf{a}, \textbf{Algorithm 1}). (\textbf{c}) Detailed scheme of organelle segmentation (\textbf{Algorithm 4}). \textbf{A} -- non-interpolating demosaicing of the RAW files of the segmented cell (input 1); \textbf{C2} -- removing undesirable objects via comparison of the cell intensity histogram with the mode of the background histogram (input 3); \textbf{C3} -- application of the binary topological envelope (input 4) to each z-stack; \textbf{B} -- creation of a binary mask via overlapping of $\omega_{\alpha,l,x,y,G} = 0$ values from two consecutive images; \textbf{C1} -- 2-D segmentation of organelles and OSFs; \textbf{D} -- 3-D reconstruction of organelles; \textbf{E} -- 3-D stacking of 2-D organelle segments. Image processing was performed in 12-bpc intensity depth and is visualized in 8 bpc.}
\end{figure}

\section{Segmentation of a cell's focal region}

\indent \indent In the first step, a cell of interest was segmented from its background by identifying green pixels whose intensities remain unchanged for each two consecutive RAW files (\textbf{Algorithm 1}, \textbf{Fig. 1b}). The intensities of the green pixels in each Bayer mask quadruplet were averaged to give quarter-resolved grayscale images,\cite{rychtarikova,tkacik} which were then subtracted. The unchanged intensities (i.e. zero values in the differential image) concurrently higher than 0 and lower than a 0.95-fold intensity mode of the cell-free second image contributed to the cumulative binary mask. In the focal region, these unchanged dark green pixels are the primary contributors to the cumulative binary mask (\textbf{Supplementary Video 1}).

This binary mask was further processed by standard morphological operations---dilating the image (a 3-px disk-shaped structuring element), filling image holes (corresponding, in the original image, to the fluorescent objects and positive light interferences in the Airy diffraction pattern,\cite{airy}) and filtering the cell of interest according to its specific features (in our case, as an object of the maximal size)---resulting in a final binary mask. The final binary mask of the cell was rescaled by a factor of two and applied to the whole z-stack of the original RAW files in order to distinguish a sum of point spread functions of the cell.

Computation of the binary mask from  RAW files' red and blue pixels did not give the desired results. Due to the high frequency of consecutive pixels with constant intensities, the image of the cell merged with its background. The reason for this may be found either in light absorption in the infra-red and ultra-violet regions\cite{niehren} or in lower photon quantum efficiency of the respective camera filters.\cite{camera} Therefore, in all segmentations, the green intensity wide range histogram was used.

The next step consists of selecting the focal sub-stack of the cell and assessing cell topography. The focal region of the z-stack was determined via clustering point information gain entropy density ($\Xi_{\alpha}$) spectra\cite{rychtarikovaB} obtained for all RAW files of the separated cell. The variable $\Xi_{\alpha}$ [bit] was derived from the R\'{e}nyi entropy as
\begin{equation}
\Xi_{\alpha,l} = \frac{1}{1-\alpha}\sum_{j=1}^{k}\log_2\frac{\sum_{j=1}^{k} p_{j,i,l}^\alpha}{\sum_{j=1}^{k} p_{j,l}^\alpha},
\end{equation}
where $p_{j}$ and $p_{j,i}$ are the probabilities of occurrence of intensity $j$ in an intensity histogram of the $l$-th image in the z-stack with and without an element of the intensity $i$, respectively. The additive term $\frac{1}{1-\alpha}\log_2\frac{\sum_{j=1}^{k} p_{j,i,l}^\alpha}{\sum_{j=1}^{k} p_{j,l}^\alpha}$ is called a point information gain ($\Gamma_{\alpha, j}$, bit) and can determine an information contribution of intensity $j$ to the intensity histogram obtained from either the whole image (a global measure $\Xi_{\alpha,Wh}$) or its part (local measures). For image processing of the presented cells, we used local values evaluated from pixels either on the vertical-horizontal cross ($\Xi_{\alpha,Cr}$) or on a 9-px circle around the examined pixel ($\Xi_{\alpha,Circle}$). The kind of local information was chosen according to the distribution of intensities in the image. Whereas the z-stacks of the MG63-\textbf{a} and L929 cells suffered from cross camera noise, the images of the MG63-\textbf{b} cell did not (\textbf{Supplementary Videos 2--3}). In the latter case, the 9-px circular type of surroundings approximately traced the borders of intracellular structures.

For the overall multifractal characterization of the images, $\Xi_{\alpha}$-spectra were calculated for a set of $\alpha = \{$0.1, 0.3, 0.5, 0.7, 0.99, 1.3, 1.5, 1.7, 2.0, 2.5, 3.0, 3.5, 4.0$\}$, for each colour channel separately. While the values $\Gamma_{\alpha,j}$, and consequently $\Xi_{\alpha,j}$, for the red and blue channels (indexed $_R$ and $_B$, respectively) were computed by eliminating one element of intensity $j$ from the respective intensity histogram, these values for the green pixels (indexed $_G$) were obtained via eliminating two elements that were relevant to the intensities of the Bayer mask quadruplet.

Matrices composed of vectors that specify each image $l$ in the z-stack via $\alpha$-dependent subvectors of the respective information context in the respective colour channel, i.e.
\begin{equation}
\Xi_{(l)} = [\Xi_{\alpha,Wh,R}, \Xi_{\alpha,Wh,G}, \Xi_{\alpha,Wh,B}, \Xi_{\alpha,Cr,R}, \Xi_{\alpha,Cr,G}, \Xi_{\alpha,Cr,B}]
\end{equation}
for series of the MG63-\textbf{a} and L929 cells and
\begin{equation}
\Xi_{(l)} = [\Xi_{\alpha,Wh,R}, \Xi_{\alpha,Wh,G}, \Xi_{\alpha,Wh,B}, \Xi_{\alpha,Cr,R}, \Xi_{\alpha,Cr,G}, \Xi_{\alpha,Cr,B},\Xi_{\alpha,Circle,R}, \Xi_{\alpha,Circle,G}, \Xi_{\alpha,Circle,B}]
\end{equation}
for the series of the MG63-\textbf{b} cell, were standardized with z-scores and underwent k-means clustering (squared Euclidean distance metric, 50 iterations) into two groups (\textbf{Algorithm 2}). Due to the spectral properties of the OSF, this clustering properly selected a focal region of the cell from the rest of the z-stack.

In \textbf{Algorithm 2}, the sub-stack of the focal region was chosen as a cluster with a RAW file whose average intensity of green pixels is the inflection point of the dependence of the average intensity of green pixels on the position of the RAW file in the z-stack. To smooth the dependence, a fourth-order polynomial was used. This part of the algorithm assumes that in the focal region the intensities over the z-stack change significantly, whereas the intensities of blurred images remain relatively constant.

The topological envelope of the cell (explained as a binary image at each z-level, \textbf{Algorithm 3}) was evaluated from the focal sub-stack of RAW files as the absolute value of the subtraction of the unblurred and blurred green pixels at the same z-level after non-interpolating de-mosaicing of green pixels of RAW files. The blurring of each particular image was performed with a filter created from a 10-px disk-shaped structural element. After that, the pixels of interest at each z-level were chosen as those brighter than twelve times the maximal intensity of the subtracted image. These pixels underwent a morphological closing (a 3-px disk-shaped structuring element), removing the undesirable pixels via morphological erosion and dilation, and computation of the binary convex hull around the rest of the binary objects. A subsequent dilation of the binary convex hull (a 20-px disk-shaped structuring element) ensured extension and rounding of the cell boundaries.

From each series, a multiplication of the number of images in the focal region by the respective average scanning step (\textbf{Table 3}) gave us a height of the part of the OSF that is occupied by the cell, i.e. 5.6, 3.6, and 5.4 $\mu$m for the MG63-\textbf{a}, MG63-\textbf{b}, and L929 cells, respectively. The shapes and the heights of the cells (\textbf{Fig. 3b} and \textbf{Supplementary Fig. 2--3b}) obtained from the bright-field microscopy images using the presented algorithm are in agreement with live cell imaging using atomic force microscopy\cite{Malakhova} (\textbf{Supplementary Fig. 4} and \textbf{Supplementary Information 1}). In the MG-63 cell line, hill-shaped cells with a protuberant nuclei, of the size of 5.2 $\pm$ 1.1 and 4.0 $\pm$ 1.0 $\mu$m on different substrates, prevail. L929 cells are approximately 0.4 $\mu$m lower and flatter. Similar results have been depicted in scanning microscopy images and described in literature.\cite{melo,jonas,chang,docheva} For the microscopy experiments, the dish bottoms were not treated.

\section{Classification, segmentation, and investigation of properties of organelles}
\indent \indent
This section describes how to extract information about the 3-D shapes and dynamics of organelles from a focal region of a z-stack of bright-field optical transmission micrographs of a detached cell. The sub-stacks of the MG63-\textbf{a}, MG63-\textbf{b}, and L929 cells were obtained with average z-step sizes of 116, 156, and 147 nm and with a scanning frequency of 0.440, 0.213, and 0.298 img. s$^{-1}$, respectively (\textbf{Table 3}).   

In order to maximize and analyze the change in the OSF's volume, we have previously derived a information-entropic variable point divergence gain\cite{rychtarikova} (PDG: $\omega_{\alpha,l,x,y,c}$, bit), which evaluates the information divergence for all pixels between two consecutive RAW files in the focal section of the z-stack:
\begin{equation}
\omega_{\alpha,l,x,y,c} = \frac{1}{1-\alpha}\log_2\frac{\sum_{j=1}^{k} p_{i,l,c}^{\alpha}}{\sum_{j=1}^{k} p_{i,(l+1),x,y,c}^{\alpha}},
\end{equation}
where $l$ is the order of an image in the focal region of the z-scan, and $x$ and $y$ are coordinates of the particular pixel in the image $l$. Probabilities $p_{i,l,c}$ and $p_{i,(l+1),x,y,c}$ describe the frequencies of occurrence of colour intensities in the image $(l)$ and in the same image after exchanging the pixel at coordinates $(x,y,l)$ for the pixel at $(x,y,(l+1))$. The $\omega_{\alpha,l,x,y,c}$-values for pixels of each colour in the RAW file's quadruplet were calculated in the same way as the $\Xi_{\alpha}$-values in \textbf{Eq. 1}: red and blue channels of the resulting quarter-resolved $\omega_{\alpha,l,x,y,R/B}$-matrices were computed after exchanging one pixel of the respective colour, whereas the green channel was obtained after exchanging two green pixels of the respective pixel quadruplet.

Compared to the simple subtraction of two consecutive images, calculating $\omega_{\alpha,l,x,y,c}$-values classifies the image pixels with respect to their probability of occurring in volume and also introduces dynamics into the examined system. Zero values of $\omega_{\alpha,l,x,y,c}$ correspond to pixels with relatively high occurrences in the image, and thus ones that do not change in a z-step. These represent stable, large, non-moving objects at a high image resolution and the smallest possible z-step, mainly organelles down to the size of one voxel. The more extremely negative or positive values of $\omega_{\alpha,l,x,y,c}$ show pixels with the highest change from image to image, which correspond mainly to moving objects. Other $\omega_{\alpha,l,x,y,c}$-values detect either sums of point spread functions of organelles or organelles themselves, which are composed of lower-occurrence intensities at the given z-level and, concurrently, whose OSFs are divergent over distances smaller than the size of the z-step.

Here, coefficient $\alpha$ represents multifractality and defines distribution. Low values of $\alpha$ merge frequently-occurring $\omega_{\alpha,l,x,y,c}$-values and separate rare pixels---the most dynamic organelles in this case. High $\alpha$ values give wider distributions of $\omega_{\alpha,l,x,y,c}$-values. A suitable value of this parameter must be always derived or estimated with regards to the multifractal character of the given intensity distribution. We decided to use $\alpha$ equal to 5 (MG63-\textbf{a}) and 6 (MG63-\textbf{b}, L929), at which value the images of the organelles' OSFs, mainly in the green channel, are adequately condensed after camera noise and another defects in the image are suppressed (\textbf{Fig. 2a}, \textbf{Fig. 3c}, and \textbf{Supplementary Figs. 2--3a}). At zero $\omega_{\alpha,l,x,y,c}$ of a higher-order $\alpha$, we already observe a strong combination of intensities of light-interferences in the image. As the size of the z-step increases, larger $\alpha$-values must be used to merge the correct image intensities.

Analysis of $\omega_{\alpha,l,x,y,c}$-values in each colour channel showed that there is mainly autofluorescence projected in the blue channel. The green channel further displays diffraction. The red channel shows also the contribution of near infra-red absorption. The application of each colour channel can be viewed when zero $\omega_{\alpha,l,x,y,c}$-values are compared with original images (\textbf{Fig. 2b}) and provide a potential for classification and recognition of organelles with the respect to their composition, without the usage of any labelling technique (cf.\cite{Eckers}).

Because computing $\omega_{\alpha,l,x,y,c}$-values for three consecutive z-stack images gives information about the shape and dynamics of organelles in the middle image, a binary mask for segmenting objects in a z-level was created by thresholding and uniting identical $\omega_{\alpha,l,x,y,c}$-values from two consecutive $\omega_{\alpha,l,x,y,c}$-matrices (input 2 in \textbf{Fig. 1c}). This mask was applied to the respective quarter-resolved image of the cell (input 1 in \textbf{Fig. 1c}), which was obtained by adapting the Bayer quadruplet's pixels of red, blue and average green to the respective colour channel. The subsequent matching of the respective binary topological mask (input 4 in \textbf{Fig. 1c}) with the image of the detached objects selected objects relevant for the given z-level (\textbf{Algorithm 4}).

The last part of the algorithm (input 3 in \textbf{Fig. 1c}) filtered irrelevant intensities from the images, which completely describe the spectral properties of the cell's image. For each colour channel, strongly light-diffracting or absorbing organelles were detached as those darker than the cell-free background. In contrast, light-emitting organelles were reconstructed from intensities brighter than the background (\textbf{Fig. 1-B1--B3}, \textbf{Algorithm 4}).

In this paper, we demonstrate a novel method for 3-D reconstruction and examination of large homogeneous non-moving cellular objects, which are projected at the most frequent value of $\omega_{\alpha,l,x,y,c}$ = 0 (\textbf{Fig. 3c} and \textbf{Supplementary Fig. 2--3c}). Apart from the large homogeneous non-moving objects (e.g. nucleoli in diffraction), the method detected objects of the size of a few voxels,\cite{lichtscheidel} which might be shown to be real objects by video-enhanced microscopy or correspond to other frequent intensities remaining constant through a z-step.

The OSFs of light-diffracting objects are substantially smaller than those of light-emitting objects, which implies that transmission microscopy has an advantage over fluorescent microscopy in biological experiments (\textbf{Supplementary Fig. 1a}). The consistently smaller number of detected objects in the green channel is probably caused either by the mathematical averaging of two green pixels of the Bayer mask quadruplet during the calculation of $\omega_{\alpha,l,x,y,c}$-values or by the broader green spectrum (caused by technical reasons, as noted above) decreasing the probability of occurrence of the same intensity between two consecutive pixels.

\begin{figure}
  \centering
  \includegraphics[width=\textwidth]{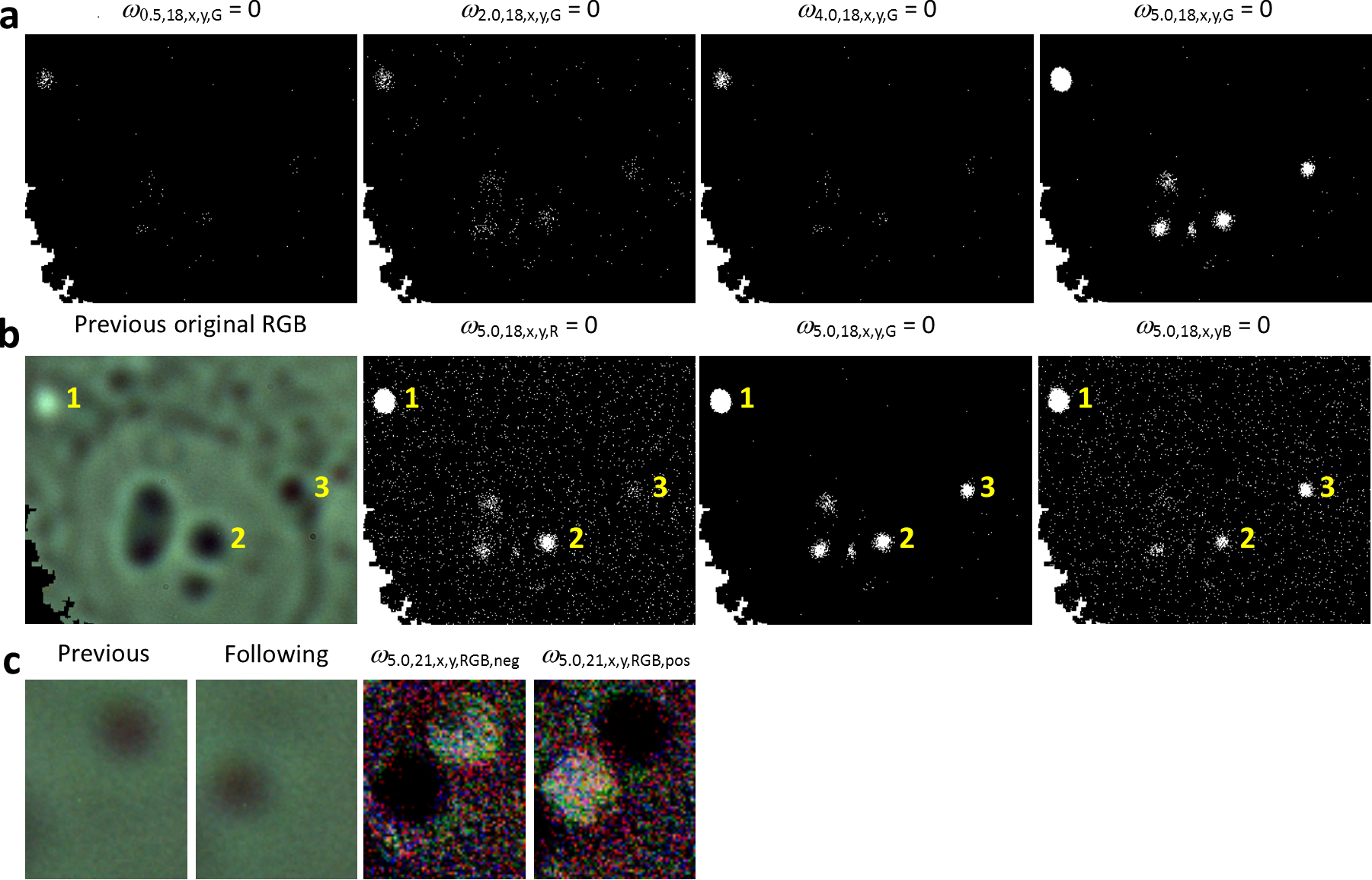}
  \caption{Details of $\omega_{\alpha,l,x,y,c}$-images of a focal plane of a z-stack of live cell from bright-field transmission optical microscopy computed using two consecutive images (illustrated on the interior of a MG63-\textbf{a} cell). (\textbf{a}) Zero values of $\omega_{\alpha,l,x,y,G}$-transformed images with points that show unchanged information at $\alpha$ equal to 0.5, 2.0, 4.0, and 5.0, respectively. The original section of the cell is identical to that of \textbf{b}. (\textbf{b}) An original RGB section of the cell (visualized in 8 bpc) and its values $\omega_{\alpha,18,x,y,c}$ = 0 for the red, green, and blue channels. Autofluorescent organelle 1 shows spectral characteristics in all colour channels. Organelle 2 (nucleolus) diffracts in the green and red channels and has weak autofluorescence due to its content of nucleolic acids. Organelle 3 bound to the nucleolar envelope is detectable only in the blue and green channels.}
  \label{Fig2.png}
\end{figure}
\begin{figure}
  \contcaption{(\textbf{c}) Movement of an organelle demonstrated on 8-bit images transformed from the original $\omega_{5.0,21,x,y,RGB}$-values in double precision floating point format (some $\omega_{5.0,l,x,y,RGB}$-values are merged into one intensity of the $\omega_{5.0,21,x,y,RGB}$-image). White and black pixels in the $\omega_{5.0,21,x,y,RGB,neg}$-image (e.g., the highest and the lowest negative $\omega_{\alpha,l,x,y,c}$-values, respectively) correspond to the position of the organelle in the previous and following original RGB images of the cell, respectively (and vice versa for the $\omega_{5.0,21,x,y,RGB,neg}$-image). The sizes of the sections in \textbf{a}--\textbf{b} is 23.732 $\times$ 19.176 $\mu$m$^{2}$ and 4.352 $\times$ 5.372 $\mu$m$^{2}$ in \textbf{c} (68 nm$^{2}$ px$^{-1}$).}
\end{figure}


\begin{figure}
  \centering
  \includegraphics[width=\textwidth]{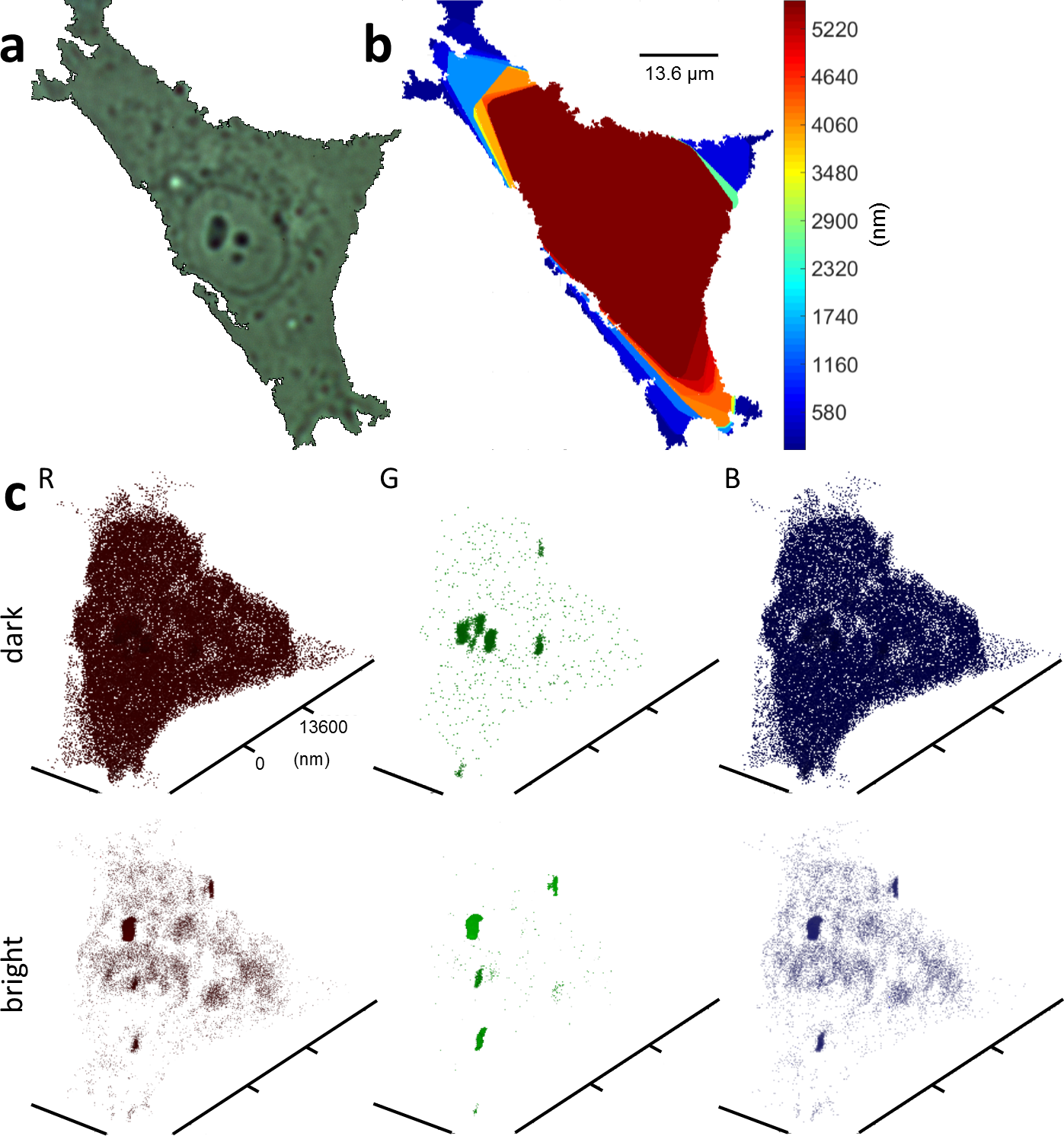}
  \caption{3-D reconstruction of a MG63-\textbf{a} cell. (\textbf{a}) An original 2-D image of the segmented MG63-\textbf{a} cell from the center of the focal region (obtained using \textbf{Algorithms 1--2} and visualized in 8 bpc). (\textbf{b}) Isocontours  of the topological space of the occurrence of the MG63-\textbf{a} cell in its OSF (calculated using \textbf{Algorithms 1--3} in Methods).}
  \label{Fig3.png}
\end{figure}
\begin{figure}
  \contcaption{(\textbf{c}) 3-D reconstruction of the large non-moving objects in the MG63-\textbf{a} interior (found using \textbf{Algorithms 1--4} in Methods). The dark objects (upper row) represent strongly light-diffracting and light-absorbing objects or pixels of destructive light interference (visualization of ranges of intensities 465--884, 1152--2169, and 593--1082 in the R, G, and B channels, respectively). The bright objects (lower row) represent autofluorescent objects or pixels of positive light interference (visualization of ranges of intensities 865--2519, 2137--3445, and 1063--3310 in the R, G, and B channels, respectively).}
\end{figure}

\section*{DISCUSSION}
\indent \indent
Knowing the distribution and mutual interactions of biomolecules can help determine the morphological and physiological state of a cell. Since the 17th century,\cite{dobell} observations of intracellular processes have been provided by microscopic techniques based on different physical principles. Imaging based on fluorescent microscopy has been a leading technique for defining the subcellular location of proteins for decades. However, fluorescent protein tagging technology suffers from some limitations, including the need for a physiological level of light-emitting protein production, mislocalization artifacts, relatively low resolution, and the necessity to intervene in the cell's physiological state after insertion of a dye.\cite{fornasiero} The breakage of the Abbe diffraction limit\cite{abbe} in fluorescent microscopy was achieved by the invention of super-resolved fluorescent imaging, which was awarded the 2014 Nobel prize in chemistry.\cite{moerner} On the other hand, contrast techniques in optical microscopy such as phase contrast,\cite{zernikeB} differential interference contrast,\cite{murphy} digital microscopic holography,\cite{kemper} interferometric microscopy,\cite{kuznetsova} and optical coherence tomography\cite{zysk} require the insertion of an optical element into the optical path of the microscope, which distorts the image of the observed biological specimen and makes image interpretation much more difficult. Electron microscopy (in both transmission and scanning modes) is an ancillary method in cell biology, since it may only be used to observe dried samples after a preparation time of several days. However, the resolution obtained by electron microscopy may go down to a few nanometers. The newest imaging method---atomic force microscopy, e.g.\cite{morris}---is a kind of non-optical topographical technique that reaches high resolution but does not provide the possibility of fully imaging intracellular composition and interactions. Connecting the benefits of these different imaging methods can be achieved by combining them; for instance, correlative light electron microscopy (CLEM, e.g.\cite{boer}) is the most well-known and commercially available example of combined imaging.

This article reports a method to comprehensively analyze the information provided by label-free bright-field photon transmission microscopy (calibrated and validated by AFM,\cite{Malakhova}), which detects minute objects of Nobelish resolution\cite{fornasiero,moerner} in a living cell. We do not develop a quantum physical theoretical foundation of the origin of information in the image. We instead follow the Extended Nijboer-Zernike Theory,\cite{braat,zernike,nijboer} which claims that the focus is at the position of the lowest/highest density of electromagnetic radiation. Provided that two points of the same energy detected by a digital camera chip lie above each other, they are considered to be a light-diffracting or light-emitting object. The extent of the detection as well as of the reliability of the interpretation is heavily limited by the microscope's optical and mechanical properties. The resolution limit is not influenced by the camera sensitivity but by the number of photons. A high number of photons enables objects to be localized (known as discriminability).\cite{lichtscheidel,urban} It is an analogy to super-resolved fluorescence microscopy, where the limit is based on a few photons.

We demonstrate some of the extraordinary properties of an image of elementary light-diffracting, light-emitting, or light-absorbing objects. Objects of the size of one camera pixel are detected. To re-phrase this observation in the terminology of the depth-of-focus in digital microscopy: the depth of focus is a step along the z-axis within which the information contained in one camera pixel remains within this pixel and is not transferred into the neighbouring pixel. Our results demonstrate that such a definition is very sharp. It means that each point in the image of $\omega_{\alpha,l,x,y,c}$ will be equal to 0. The fact that we have observed only a few points at $\omega_{\alpha,l,x,y,c}$ = 0 indicates that objects' spread functions, which give rise to the image in these camera points, have homogeneous intensity over more than one z-level. The latter conclusion indicates that objects detected with $\omega_{\alpha,l,x,y,c}$ = 0 at all $\alpha$ values are located within the volume of the voxel. For these objects, the information obtained by our approach is equivalent to a 3-D reconstruction constructed from electron microscopy images. The detection limit of other objects, which gives rise to a certain distortion in the optical paths, is solely technical. It is due to mechanical precision in the z-step and x-y reproducibility, the size of the camera pixel, the objective magnification, a simple optical path, homogeneous illumination, the scanning frequency, the distribution of camera noise, the bit depth of the camera, and image storage and computational capacity.

\section*{METHODS}

\textbf{Cell cultivation}


\noindent{MG-63 (human osteosarcoma, Serva, cat. No. 86051601) and L929 (mouse fibroblast, Serva, cat. No. 85011425) cell lines were grown at low optical density overnight at 37$^\circ$C in a synthetic dropout media with 30\% raffinose as the sole carbon source. The nutrient solution for the MG-63 cells consisted of: 86\% EMEM, 10\% newborn-calf serum, 1\% antibiotics and antimycotics, 1\% L-glutamine, 1\% non-essential amino acids, 1\% $\mbox{NaHCO}_{3}$ (all components were purchased from PAA Laboratories). During microscopy experiments, cells were cultivated in a Bioptech FCS2 Closed Chamber System.}

\noindent{\textbf{Microscopy}}

\noindent{Microscopy of a living MG-63 cell culture was performed using a versatile sub-microscope: a nanoscope developed for the Institute of Complex Systems FFPW by the company Optax Ltd. (Czech Republic). The optical path consisted of two Luminus 360 light emitting diodes, a condenser system, a firm sample holder, and an 40$\times$ objective system made of two complementary lenses that allow a change of distance between the objective lens and the sample. The UV and IR light was blocked by a 450-nm long-pass filter and a 775-nm short-pass filter (Edmund Optics), respectively. Next, a projective lens magnified the image onto a Kodak KAI-16000 camera chip with 4872 $\times$ 3248 resolution and 12-bit colour depth. The size of the original camera pixel using primary magnification was 34 $\times$ 34 nm$^2$. The z-scan was performed automatically by a programmable piezomechanic (servo) motor. The scanning conditions are presented in \textbf{Table 1} and \textbf{Supplementary Data 1}.}

\noindent{\textbf{Image processing algorithm}}

\noindent{The relevant stacks of micrographs (ca. 2/3 of the original z-stack) were selected from the original z-stacks using the "ILCZ" (MG63-\textbf{b}, L929) tag from the Exif metadata of each image using the file pngparser.exe (in imagesInfo.txt in \textbf{Supplementary Material} available via ftp connection\cite{ftp}). For the MG63-\textbf{a} cell, the same process was performed using Matlab$^{\textregistered}$ scripts: RelImgSelection.m and Shift.m (for image alignment). The average steps and total scanning times are described in \textbf{Table 1}.}

The bulk of the image processing and analysis of the bright-field optical micrographs were carried out with Matlab$^{\textregistered}$ R2014b software fortified by Image Processing and Statistics Toolboxes (Mathworks, USA) using an OrganelleExtraction script package (ICS FFPW, USB, Czech Republic). The variables Point Information Gain Entropy Density ($\Xi_{\alpha}$, PIED) and Point Divergence Gain ($\omega_{\alpha,l,x,y,c}$, PDG) (\textbf{Eqs. 1} and \textbf{3}) were computed using Image Info Extractor Professional v.b11 software (ICS FFPW, USB, Czech Republic; a GBRG Bayer grid) and stored in double precision floating point format in Matlab$^{\textregistered}$ structure arrays. The differences in image processing of the cells are shown in \textbf{Table 2}. The basic algorithms for segmentation of cells and intracellular objects are written below. The optimized m-files, software, and original and processed data are available via ftp connection\cite{ftp}.

\begin{table}
\footnotesize
\caption{Microscope Setup}
\label{Tab1}
\begin{center}
\begin{tabular}{l l l l l l l l p{3cm}}
\hline
    \bfseries Cell & \bfseries Series & \bfseries  & \bfseries & \bfseries Camera & \bfseries & \bfseries  & \bfseries Piezo\footnote{If yes, the image series underwent image alignment (registration).}\\
    \hline
    \bfseries & \bfseries Number of img.\footnote{The original number of image in the series before z-step selection is parenthesized.} & \bfseries Step (nm) & \bfseries Time  (min:s) & \bfseries Offset & \bfseries Gain & \bfseries Exposure (ms) & \bfseries\\
    \hline
    MG63-\textbf{a} & 93 (155) & 119 & 03:35.4 & 0 & 268 & 3327 & Yes \\
    MG63-\textbf{b} & 128 (201) & 150 & 10:22.7 & 266 & 347 & 2466 & No\\
    L929 & 173 (358) & 158 & 11:09.0 & 221 & 336 & 2632 & No\\
\hline
\end{tabular}
\end{center}
\end{table}
\newpage

\begin{table}
\footnotesize
\caption{Image processing of the presented cells}
\label{Tab2}
\begin{center}
\begin{tabular}{l l l l l l p{3cm}}
\hline
    \bfseries Cell & \bfseries Coordinates of background & \bfseries Selection of focus & \bfseries 3-D imaging &  \\
    \hline
    \bfseries & \bfseries $x_1$, $x_2$, $y_1$, $y_2$ & \bfseries Local $\Xi_\alpha$ & \bfseries $\alpha$ for $\omega_{\alpha,l,x,y,c}$ & \bfseries R, G, B threshold \\
    \hline
    MG63-\textbf{a} & 4, 268, 652, 894 & cross & 5 &  \\
    MG63-\textbf{b} & 26, 322, 1296, 1618 & cross, 9-px circle & 6 & 1250, 2300, 1500 \\
    L929 & 144, 792, 803, 1268 & cross & 6 & 1000, 1700, 1170 \\
\hline
\end{tabular}
\end{center}
\end{table}

\begin{table}
\footnotesize
\caption{Characterization of the focal regions}
\label{Tab3}
\begin{center}
\begin{tabular}{l l l l l l l l p{3cm}}
\hline
    \bfseries Cell & \bfseries Coordinates of position  & \bfseries Number of img. & \bfseries Average step & \bfseries z-Height & \bfseries Time & \bfseries Img. frequency \\
\bfseries  & \bfseries $x_1$, $x_2$, $y_1$, $y_2$  & \bfseries & \bfseries (nm) & \bfseries (nm) & \bfseries (min:s) & \bfseries (s$^{-1}$) \\
    \hline
    MG63-\textbf{a} & 55, 1928, 1, 2278 & 49 & 116 & 5568 & 1:51.3 & 0.440 \\
    MG63-\textbf{b}  & 1, 2436, 139, 3248 & 24 & 156 & 3588 & 1:52.2 & 0.213 \\
    L929 & 767, 1798, 341, 1432 & 38 & 147 & 5436 & 2:07.7 & 0.298 \\
\hline
\end{tabular}
\end{center}
\end{table}
\newpage

\lstset{language=Matlab,%
    basicstyle=\footnotesize,
    breaklines=true,%
    morekeywords={matlab2tikz},
    keywordstyle=\color{black},%
    morekeywords=[2]{1}, keywordstyle=[2]{\color{black}},
    identifierstyle=\color{black},%
    commentstyle= \textit,%
    showstringspaces=false,
    numberstyle={\tiny \color{black}},
    numbersep=9pt, 
    emph=[1]{for, end, if, INPUT, OUTPUT, raw1, raw2, G1, G2, BM, zeros, darkZeroG, zeroG, difG, CellBM, cumG, pied, idx, focReg, averInt, rawCell, G, fitInt, derInt, rawCell2, PDG1C, PDG2C, dark1, bright1, bg1, Cell2, PDG1C0, PDG2C0, OrgBM, envelope2, Cell2dark, Cell2bright, n, x1, x2, y1, y2, m, c1, c2, dev, inflex, idxInflex, Cell, filtCell, difCell, cutDifCell, thresh, c, thCutDifCell, closeObjects, bigObjects, envelope, level},emphstyle=[1]\textbf, 
}

\newpage
\noindent \textbf{Algorithm 1: creating a binary mask to segment a cell of interest from a bright-field optical transmission z-stack}
\begin{lstlisting}
--------------------
INPUT:
      n RAW files of with cell of interest;
      x1, x2, y1, y2 as coordinates of the background region;
      c1 as a threshold constant of the background (c1 = 0.95);
      c2 as a size of the structural image dilating element (c2 = 3);
      BM as a zero matrix of the quater size than the RAW file
OUTPUT:
      CellBM as a binary mask of the cell of interest
--------------------
for i = 1:(n-1)
    raw1 = readIm(i);
    raw2 = readIm(i+1);
          % read the (i)th and (i+1)th RAW file, respectively
    
    G1 = demosaicG(raw1);
    G2 = demosaicG(raw2);
          % create a quater-resolved image by averaging two green pixels of each Bayer mask's quadruplet in the (i)th and (i+1)th RAW file, respectively
    m = findMode(G2(x1:x2, y1:y2));
          % find the intensity mode of the background in the (i+1)th image

    difG = double(G2) - double(G1);
          % subtract G-channels of two consecutive images
    zeros = (difG == 0);
          % find intensities of the subtractive image equals to 0
    zeroG = G2 .* uint(zeros); 	% select unchanged G-intensities from G-image
    darkZeroG = (zeroG < c1*m) & (zeroG > 0);
          % find unchanged G-intensities darker than the c1% of the value of the mode of background and brighter than zero intensity
    cumG = cumG + darkZeroG;
          % calculate a cumulative binary mask from dark unchanged green intensities
end

BM = (cumG > 0);	% threshold non-black pixels in binary mask
BM = dilateBW(BM, strel('disk', c2));
      % dilate the binary mask with the c2-px sized disk structural element
BM = fillHolesBW(BM);	% fill holes in the binary mask
CellBM = filterCell(BM);
      % filter a cell of interest from the binary mask
CellBM = resizeIm(CellBM, 2);
      % increase (2x) the size of the binary mask to achieve the size of the original RAW files
--------------------
\end{lstlisting}

\newpage
\noindent \textbf{Algorithm 2: selecting the focal region using $\Xi_{\alpha}$ values}
\begin{lstlisting}
--------------------
INPUT:
      n RAW files of cell of interest;
      pied as a matrix of the size of (number of colour image channels x number of alpha) x n
OUTPUT:
      focReg as a matrix specifying images which belong to the focal region
--------------------
pied = zscore(pied);
      % calculate a z-score for each sample (image) spectrum over alphas
idx = cluster(pied, 2);
      % cluster samples (images) into 2 groups (via k-means algorithm with Euclidian distance) and assign a number of group to each sample (image) into vector idx

      % find a focal plane of the series as the image of the average darkest intensity
averInt = zero(n, 1);
      % create a zero matrix averInt for the storage of RAW files' average G-intensities

for i = 1:n
    rawCell = readIm(i);
    G = demosaicG(rawCell);
    averInt(i) = nonzeroMean(G);
end
      % calculate the average intensity of G-channel for each image in the z-stack (omit a black background from the calculation)

fitInt = smoothCurve(averInt);
      % fit the dependency of average G-intensity on the position in the z-stack by a smooth curve (a polynom of order 4)
inflex = findInflexion(fitInt);	
      % find an inflexion point of the smoothed curve, which corresponds to the best focused image in the z-stack
idxInflex = findIdx(idx == inflex);
      % find the number of cluster in the idx matrix, which corresponds to the image at the position of the inflexion point
focReg = findRegion(idx == idxInflex);
      % find the focal region as the number of cluster with the image, which corresponds to the inflexion point of the intensity curve
--------------------
\end{lstlisting}

\newpage
\noindent \textbf{Algorithm 3: obtaining the topography}
\begin{lstlisting}
--------------------
INPUT:
      n RAW files of a cell of interest;
      c as a threshold constant (c = 12)
OUTPUT:
      envelope as a binary matrix with a levels of the topological set of the cell
--------------------
BM = resizeIm(logical(readIm(1)), 0.5);
      % create a binary image of the cell, which is of a quater resolution in comparison to the original image
envelope = zeros(size(BM, 1), size(BM, 2), length(flr));
      % create a zero matrix of the size of the whole z-stack
envelope(:,:,2) = BM;
      % save the binary mask of the original cell into the second layer of the envelope matrix

BM = erodeBW(BM, strel('disk', 10));
      % erode the original binary mask with a structural element (a 10-px disk) to remove the edges of the cell in a image obtained via a subtraction of the blurred and unblurred input image
BM([1:10, end-10:end], :) = 0;
BM(:, [1:10, end-10:end]) = 0;
       % remove the edges of the bw-image which touch the edges of the picture

for i = 3:n
    rawCell = readIm(i);
    Cell = demosaicG(rawCell); % demosaic a G-channel

    filtCell = filterIm(Cell);
          % filter the image Cell with a circular averaging filter (pillbox of the 10-px radius) to create a blurred image of the cell
    difCell = abs(double(Cell) - double(filtCell));
          % calculate absolute values in the image, which is a subtraction of a focal image and its blurred version 
    cutDifCell = difCell .* double(BM);
          % cut the edges of the cell using a binary mask BM;

    thresh = findMax(cutDifCell)/c;
          % calculate a threshold for the selection of the immovable objects as a ratio of the maximal value in the subtractive image cutDifCell and input constant c
    thCutDifCell = cutDifCell > thresh;
          % threshold the values in the subtractive image higher than the threshold thresh
    closeObjects = closeBW(thCutDifCell);
          % perform image closing (with a 3px disk structural element on the threshold image)
    bigObjects = filterObjects(closeObjects);
          % remove small objects in the closed image
    envelope(:,:,i) = uniteObject(bigObjects);
          % unite the rest of objects and create an envelope
    envelope(:,:,i) = envelope(:,:,i) .* envelope(:,:,i-1);
          % apply the previous envelope to the current one to make the mask gradually smaller
end
--------------------
\end{lstlisting}

\newpage
\noindent \textbf{Algorithm 4: 2-D segmentation of objects}
\begin{lstlisting}
--------------------
INPUT:
      rawCell2 as the second RAW file of a cell of interest from two consecutive images;
      PDG1C and PDG2C as matrices of point divergence gain values calculated for the respective colour channel of two consecutive images, respectively;
      m as an intensity mode of the background in the respective colour channel of the second image;
      envelope2 as a binary envelope of the cell at the second z-level
      level as a value of PIED (level = 0 for large non-moving objects)
OUTPUT:
      dark1 and bright1 as 2D segments of organelles of different spectral properties
--------------------
Cell2 = demosaicC(rawCell2);
      % create a quater-resolved image of the cell in its respective colour channel via non-interpolating algorithm\cite{Tkacik}

PDG1C0 = PDG1C == level;
PDG2C0 = PDG2C == level;
      % in each PDG matrix, threshold values level
OrgBM = (PDG1C0 + PDG2C0) > 0;
      % from positive values in the summed binary images with thresholded levels, create a binary mask for the export of objects organelles

Cell2bright = Cell2;	% duplicate the matrix Cell2 

Cell2(Cell2 >= m) = 0;
      % select autofluorescent objects (and positive interferences)
Cell2bright(Cell2bright <= m) = 0;
      % select diffraction and absorption (and negative interferences)

dark1 = (Cell2 .* uint(OrgBM)) .* uint(envelope2);
bright1  = (Cell2bright .* uint(OrgBM)) .* uint(envelope2);
      % apply the binary mask with organelles and that with the topology to the debayerized image of the cell
--------------------
\end{lstlisting}

\section*{SUPPLEMENTARY INFORMATION}

\textbf{Supplementary Information 1.} 3-D shapes and heights of MG63 and L929 cells obtained using an atomic force microscope Axio Observer.A1, Zeiss in contact mode.

\noindent{\textbf{Supplementary Figure 1} (\textbf{a}) \textit{left} -- The Extended Nijboer Zernike simulation of fluorescence (parameters NA = 0.5, d = 0.2 $\mu$m, $\lambda$ = 0.2$\mu$m, m = 0, n = 0). \textit{right} -- A real (measured) object spread function of a 0.22 $\mu$m bead in diffraction with sections of RGB images. The central sections of object spread functions show the positions of focus. (\textbf{b}) A model of phenomena of geometric optics that occur during the interaction of light with an object. The main process is diffraction. In the case of total diffraction of light at the sample interface, it can be considered that the intensities of the sample interior are black and constant, whereas the intensities of light interferences around the sample are brighter and change more in space.}

\noindent{\textbf{Supplementary Figure 2}. 3-D reconstruction of a MG63-\textbf{b} cell. (\textbf{a}) An original 2-D image of the segmented MG63-\textbf{b} cell from the center of the focal region (obtained using \textbf{Algorithms 1--2} and visualized in 8 bpc). (\textbf{b}) Isocontours  of the topological space of the occurrence of the MG63-\textbf{b} cell in its OSF (calculated using \textbf{Algorithms 1--3} in Methods). (\textbf{c}) 3-D reconstruction of the large non-moving objects in the MG63-\textbf{b} interior (found using \textbf{Algorithms 1--4} in Methods). The dark objects (upper row) represent strongly light-diffracting and light-absorbing objects or pixels of destructive light interference (visualization of ranges of intensities 647--921, 1216--1741, and 747--1030 in the R, G, and B channels, respectively). The bright objects (lower row) represent autofluorescent objects or pixels of positive light interference (visualization of ranges of intensities 910--1475, 1723--2584, and 1026--1495 in the R, G, and B channels, respectively).}

\noindent{\textbf{Supplementary Figure 3}. 3-D reconstruction of a L929 cell. (\textbf{a}) An original 2-D image of the segmented L929 cell from the center of the focal region (obtained using \textbf{Algorithms 1--2} and visualized in 8 bpc). (\textbf{b}) Isocontours  of the topological space of the occurrence of the MG63-\textbf{b} cell in its OSF (calculated using \textbf{Algorithms 1--3} in Methods). (\textbf{c}) 3-D reconstruction of the large non-moving objects in the L929 interior (found using \textbf{Algorithms 1--4} in Methods). The dark objects (upper row) represent strongly light-diffracting and light-absorbing objects or pixels of destructive light interference (visualization of ranges of intensities 445--763, 676--1257, and 533--920 in the R, G, and B channels, respectively). The bright objects (lower row) represent autofluorescent objects or pixels of positive light interference (visualization of ranges of intensities 757--1102, 1247--1630, and 908--1212 in the R, G, and B channels, respectively).}

\noindent{\textbf{Supplementary Figure 4.} Live cell imaging using an atomic force microscope Axio Observer.A1, Zeiss in contact mode. (\textbf{a}) 3-D images and heights of a MG63 (similar to presented cells MG63-\textbf{b} and L929). (\textbf{b}) Average size of MG63 and L929 cells spreading on a mat coated with either fibrinogen or fibronectin. The standard deviations were calculated from 8 cells for the MG63 cell line on both substrates, 6 cells for the L929 cell line on fibrinogen, and 3 cells for the L929 cell line on fibronectin.}

\noindent{\textbf{Supplementary Data 1.} Image pre-processing of bright-field transmission z-stacks, including information about the positions of images in the z-stacks. The gray sections correspond to the focal regions. Average values of z-positions and scanning times are reported in Tables 2--3.}

\noindent{\textbf{Supplementary Video 1.} The creation of a binary mask for segmentation of cells over the whole z-stack of 12-bit RAW files from bright-field optical transmission (described in \textbf{Algorithm 1}, demonstrated on the MG63-\textbf{a} cell). The white points correspond to the zeros in a differential image calculated from the dark green pixels of two consecutive images. With an increasing number of z-levels, white points gradually accumulate in the binary image. The highest amount of these points is achieved in the focal region (z-levels 36--84). After passing the algorithm through the whole z-stack, the binary image underwent the morphological operations of dilation, filling holes, and filtering cells.}

\noindent{\textbf{Supplementary Video 2} The points of unchanged intensity between two consecutive images in the focal region of the z-stack of 12-bit RAW files from bright-field optical transmission of the MG63-\textbf{a} cell. The white points were found by overlapping two differential images calculated from the green channels of three consecutive images (instead of $\omega_{\alpha,l,x,y,c}$ = 0 in \textbf{Algorithm 4}). Without merging levels of similar intensities in histograms of original images due to the R\'{e}nyi entropy, no organelles were detected. The algorithm only highlighted the cross camera noise (of dark green intensities, cf. \textbf{Supplementary Video 2}). The course of the video for the MG63-\textbf{b} cell was similar.}

\noindent{\textbf{Supplementary Video 3} The points of unchanged intensity between two consecutive images in the focal region of the z-stack of 12-bit RAW files from bright-field optical transmission of the L929 cell. The points were found by overlapping two differential images calculated from the dark green pixels of three consecutive images (instead of $\omega_{\alpha,l,x,y,c}$ = 0 in \textbf{Algorithm 4}). Since the z-stack of images is noise-free, some organelles were already detected via simple subtraction of consecutive images (cf. \textbf{Supplementary Video 3}).}

The image data, Matlab$^{\textregistered}$ codes, and other software are available via ftp connection.\cite{ftp}



\begin{addendum}
\item[ACKNOWLEDGEMENTS] This work was financially supported by CENAKVA CZ.1.05/2.1.00/01.0024 and The CENAKVA Centre Development CZ.1.05/2.1.00/19.0380. Authors thank to Vladim\'{i}r Kotal and Dalibor \v{S}tys Jr. for their contribution to instrument and instrument software testing, and Hana N\'{a}varov\'{a}, Monika Homolkov\'{a}, \v{S}\'{a}rka Beranov\'{a}, Pavl\'{i}na Tl\'{a}skalov\'{a} and students of Summer Schools in Nov\'{e} Hrady (Magdalena Koutov\'{a}, Marie Hyblov\'{a}, Adam Charv\'{a}t, Marco Gom\'{e}z, Kateryna Akulich, Kaijia Tian, Isabel Mogollon and others) for laboratory work.

\item[AUTHOR CONTRIBUTION]

D.S. conceived of the project, designed the experimental device and tested its performance, R.R. and T.N. analyzed the data and developed the segmentation algorithm based on the first version of algorithm by K.S., P.M. developed entropy-based computing software, J.U. derived the entropy-based variables, R.R. and R.S. scanned the cells and tested the algorithm and device software, T.M. measured standard samples of beads to verify the algorithm, D.M. performed the AFM measurements of live cells, R.R. prepared a manuscript, and all authors discussed and contributed to the final version.

\item[COMPETING FINANCIAL INTERESTS]

The authors declare that they have no competing financial interests.

\item[Correspondence]

Correspondence and requests for materials should be addressed to R.R. or D.S.\linebreak(emails: rrychtarikova@frov.jcu.cz, stys@jcu.cz).
\end{addendum}


\end{document}